\documentclass[]{JHEP}
\usepackage{citesort,graphicx,subfigure,axodraw}
\usepackage{amsmath, amssymb, bm}
\renewcommand\vec[1]{\mathbf{#1}}
\newlength{\symbwidth}
\newlength{\owidth}
\newcommand{\hatfrac}[2]{\frac{\hat{#1}}{\hat{#2}}}

\newcommand{\tr}{{\mathrm{tr}}}

\newcommand{\be}{\begin{equation}}
\newcommand{\ee}{\end{equation}}
\newcommand{\bea}{\begin{eqnarray}}
\newcommand{\eea}{\end{eqnarray}}

\newcommand{\bear}{\begin{array}{l}}
\newcommand{\eear}{\end{array}}

\newcommand{\ie}{{\it i.e.}\ }
\newcommand{\cf}{{\it cf.}\ }
\newcommand{\eg}{{\it e.g.}\ }

\newcommand{\aka}{a.k.a.\ }

\newcommand{\viz}{{\it viz.}\ }
\newcommand{\half}{\frac{1}{2}}

\def\eq#1{(\ref{#1})}
\def\eqs#1#2{(\ref{#1},\ref{#2})}
\def\sec#1{sec.\ \ref{#1}}
\def\fig#1{fig.\ \ref{#1}}
\def\ins#1#2#3{\hskip #1cm \hbox{#3}\hskip #2cm}

\def\phi{ \varphi }
\def\p{{\bf p}}

\def\A{{\bar A}}
\def\B{{\bar B}}
\def\b#1{{\bar#1}}
\def\mi{\!-\!}
\def\pl{\!+\!}

\title{Structure of the MHV-rules Lagrangian}

\author{James H. Ettle and Tim R. Morris\\
School of Physics and Astronomy, University of
Southampton,
Highfield, Southampton SO17 1BJ, U.K.\\ E-mails:
\email{jhe@phys.soton.ac.uk}, \email{T.R.Morris@soton.ac.uk}}

\maketitle

\preprint{SHEP 06-17}

\abstract{Recently, a canonical change of field variables was
proposed that converts the Yang-Mills Lagrangian into an MHV-rules
Lagrangian, \ie one whose tree level Feynman diagram expansion
generates CSW rules. We solve the relations defining the canonical
transformation, to all orders of expansion in the new fields,
yielding simple explicit holomorphic expressions for the expansion
coefficients. We use these to confirm explicitly that the three,
four and five point vertices are proportional to MHV amplitudes
with the correct coefficient, as expected. We point out several
consequences of this framework, and initiate a study of its
implications for MHV rules at the quantum level. In particular, we
investigate the wavefunction matching factors implied by the
Equivalence Theorem at one loop, and show that they may be taken
to vanish in dimensional regularisation.}

\begin{document}
\section{Introduction}
\label{Introduction}

The standard approach to computing perturbative scattering
amplitudes is to develop a Feynman diagram expansion using Feynman
rules, which in turn follow directly from a bare Lagrangian. While
this procedure is very well understood \cite{books}, the
complexity of the calculation grows so rapidly with increasing
order that it seriously challenges our ability, especially at one
loop and higher, to compute background processes to the accuracy
that must be known if we are to exploit fully the new physics
potential of the LHC \cite{exp}.

Stemming from two remarkable papers \cite{Witten1,CSW}, dramatically simpler
methods have been developed which could help solve this problem. These so far
apply mostly to scattering amplitudes in gauge theories at tree level
\cite{tree,bcftree} but also to some one-loop amplitudes
\cite{globRisager:19-22,oneloop,Bern:2004bt,bcfoneloop}.

The starting point is the set of maximally helicity-violating (MHV) amplitudes:
the tree-level colour-ordered partial amplitudes for $n_-=2$ negative helicity
gluons and any number $n_+\ge1$ of positive helicity gluons.\footnote{Throughout
the paper we label all external momenta as out-going.} Despite the factorial
growth in complexity of the underlying Feynman diagrams for increasing $n_+$,
when these are written in terms of some associated two-component spinors, the
amplitudes collapse to a single simple ratio \cite{ptbg}. As if this were not
astonishing enough, considerations of topological string theory in twistor space
\cite{Witten1} led Cachazo, Svrcek and Witten to conjecture that arbitrary
($n_+,n_-$) tree-level amplitudes of gluons may be calculated by sowing together
certain off-shell continuations of these MHV amplitudes with scalar propagators,
using colour-ordered Feynman rules \cite{CSW}. These ``MHV-rules'' (\aka CSW
rules) result in much simpler expressions for generic small $n_-$, growing in
complexity only polynomially with increasing $n_+$. Under parity, we can
exchange  $n_+\leftrightarrow n_-$, resulting in an alternative expansion via
${\overline{\rm MHV}}$-rules.

The MHV rules were proven indirectly as a consequence of another development
\cite{Britto:2005fq}: the BCFW recursion, an expansion of colour-ordered
amplitudes involving simultaneously both MHV and ${\overline{\rm MHV}}$
sub-amplitudes, which in some cases leads to even more compact expressions
albeit at the expense of introducing unphysical poles in intermediate terms. The
recursion equation results from using Cauchy's theorem on a carefully chosen
complex continuation, to reconstruct the amplitude from its poles. This idea has
been generalised to provide a direct proof of the MHV rules
\cite{Risager:2005vk}, and applied and extended at tree-level to both Yang-Mills
and other theories \cite{bcftree}. It has also been generalised to one-loop
amplitudes \cite{bcfoneloop,Bern:2004bt}, although  the appearance of physical
cuts, spurious cuts, higher poles in complex momenta, and the need for
regularisation in general, limits the power of this idea.

In a separate and initially unexpected development, the MHV rules have been
applied successfully at one loop \cite{globRisager:19-22,Bill,Bern:2004bt} by
again tying together the same off-shell continuation of the MHV amplitudes with
scalar propagators. This is meant to provide the cut-constructible parts of
one-loop amplitudes (which is the whole one-loop amplitude in theories with
unbroken supersymmetry). Although much evidence supports this contention, a full
proof has been missing \cite{Bill2} --- until now (see below).

The cut-constructible parts of one-loop amplitudes are specified
because these are directly related to tree amplitudes via their
cuts. No claim is made therefore to generate the full one-loop
amplitude in non-supersymmetric theories from MHV rules. Indeed,
it is known that certain non-constructible (parts of) one-loop
amplitudes are {\sl not generated} by MHV rules \cite{CSW2}, a
fact that we will return to in the conclusions.

As can be gleaned even from this short survey, a feature of these new
developments is that they lie outside the Lagrangian framework, proceeding by a
combination of inspired conjecture and varying levels of proof.

All this potentially changes with Mansfield's paper however
\cite{Paul}. According to this paper, a change of field variables
satisfying certain specific properties transforms the standard
Lagrangian into an equivalent MHV-rules Lagrangian, \ie one
involving effectively a complex scalar whose vertices are
proportional to the CSW off-shell continuation of the MHV
amplitudes, so that it directly generates at tree level the CSW
rules. (For a different approach to such a Lagrangian, see
ref.~\cite{Gorsky:2005sf}. For approaches via (ambi)twistor space
see ref.~\cite{lag2}.)

Assuming the validity of this proposal, we can apply at once the
well developed quantum field theory framework \cite{books} to
confirm and extend these methods. For example, if we side-step for
the moment the issue of regularisation, by concentrating on the
cut-constructible parts of one-loop amplitudes at non-exceptional
momenta, the existence of the MHV-rules Lagrangian shows
immediately that the program of
refs.~\cite{globRisager:19-22,Bill,Bern:2004bt,Bill2} is correct
because the application of MHV rules at one loop simply means in
this framework that one constructs the one loop amplitudes by
using this equivalent Lagrangian. (Actually, we need to take into
account `ET matching factors', certain wave function
renormalization matching factors that arise from applying the
equivalence theorem \cite{books,ET}, however we will see later
that they vanish.)

It also means that, up to modifications which are necessary to provide a full
regularisation (which is also much easier to figure out within standard quantum
field theory) and the ET matching factors, the MHV rules must work for the full
amplitudes at any number of loops. In particular we can expect that MHV rules
supply the full amplitude at any loop in finite supersymmetric theories (\eg
${\cal N}=4$ Yang-Mills).

We make further comments about this framework in the conclusions.

The structure of the rest of the paper is as follows. In the next
section we briefly review Mansfield's construction \cite{Paul},
which is based on light-cone quantization, and state precisely the
form the MHV-rules Lagrangian should take, paying attention to
conventions. We note in passing that the arbitrary null vector
introduced by Cachazo, Svrcek and Witten to define off-shell
continuations is exactly the one defining light-cone time.

In \sec{Expansion}, we give the transformation explicitly to all
orders by deriving recurrence relations, which we then solve. We
find that the expansion coefficients take simple forms and are in
particular already holomorphic. From Mansfield's arguments it
follows that the resulting vertices yield the MHV-rules
Lagrangian, however we nevertheless check explicitly that the
required three, four and five point vertices are obtained.

In \sec{ET}, we investigate the one-loop ET matching factors.
These are examples of terms that could not be anticipated using
the earlier methods. In the massless case however (as here) it is
easy to argue that they vanish in dimensional regularisation.
Although we are missing the full regularisation, the leading
divergent pieces should be able to be trusted. We check explicitly
and find indeed that they are forced to vanish.

Finally, in section \sec{Conclusions}, we draw our conclusions.

\section{Notation and Transformation}
\label{Transformation}

In this section we review the form of the transformation to an MHV-rules
Lagrangian as proposed in ref.~\cite{Paul}. We define closely allied notation,
but pin down several factors involved in comparison with the MHV rules.

\subsection{Preliminaries}

Mansfield maps from the Minkowski coordinates $(t,x^1,x^2,x^3)$
using a $(+,-,-,-)$ signature metric to ones appropriate for
light-front quantization \ie quantization surfaces of constant
$x^0=\mu\cdot x$, where $\mu^\nu$ is some constant null-vector.
Defining Minkowski coordinates so that $\mu^\nu = (1,0,0,1)$, the
map is
\be
x^0=t-x^3,\quad x^{\bar 0}=t+x^3,\quad z=x^1+ix^2, \quad {\bar
z}=x^1-ix^2.
\label{coord}
\ee
In these coordinates, the metric has covariant components $g_{0{\bar
0}}=g_{{\bar 0}0}=-g_{z\bar z}=-g_{\bar z z}=1/2$, all others
being zero.

We will mostly deal with covariant vectors (1-forms) for which it
is useful to introduce a more compact notation, thus we write
$(p_0, p_{\bar 0}, p_z, p_{\bar z}) \equiv (\check{p}, \hat p, p,
\bar p)$. This allows us, in sympathy with the literature, to
write components of external momenta  simply by the number of the
leg with the appropriate decoration, thus the $n^{\rm th}$
momentum $p^\mu_n$ will simply be written as $(\check{n}, \hat n,
\tilde n, \bar n)$. Note that we put a tilde over the $z$
component in this case, so that $\tilde n$ will not be confused
with a numerical factor of $n$ [see \eg
\eqs{eq:anglebracket-mom}{bracket-mom}].

We can write any 4-vector in the form of a bispinor with
components $p_{\alpha\dot \alpha}$ as
\be
\label{quaternion}
\left(p_{\alpha\dot \alpha}\right) = \left(p^t\:\delta_{\alpha\dot
\alpha} + \vec{p} \cdot \bm{\sigma}_{\alpha\dot \alpha}\right) = 2
\begin{pmatrix}
    {\check p} & -  p \\
    -\bar p & \hat p
\end{pmatrix},
\ee
where $\bm{\sigma}$ is the usual 3-vector formed of the Pauli
matrices. If $p^\mu$ is null, $\check{p} \hat p =   p \bar p$ and
the bispinor factorises: $p_{\alpha\dot \alpha} = \lambda_\alpha
\tilde\lambda_{\dot \alpha}$. For real $(p^t,\vec{p})$,
$\lambda_\alpha$ and $\tilde\lambda_{\dot \alpha}$ are related by
complex conjugation. For momenta, it is helpful to make the
choice:
\begin{equation}
\label{eq:spinor-choice}
\lambda = \sqrt 2 \begin{pmatrix}
    -{  p}/{\sqrt{\hat p}} \\
    \sqrt{\hat p}
    \end{pmatrix}
\quad\text{and}\quad \tilde\lambda = \sqrt 2
\begin{pmatrix}
    -{\bar p}/{\sqrt{\hat p}} \\
    \sqrt{\hat p}
    \end{pmatrix}.
\end{equation}
Of particular importance for MHV rules is the `angle bracket'
invariant, which we can now express as
\begin{equation}
\label{eq:anglebracket-mom}
\langle 1\:2 \rangle
    := \epsilon^{\alpha\beta}\lambda_{1\alpha} \lambda_{2\beta}
    = 2 \frac{(1\:2)}{\sqrt{\hat 1 \hat
    2}},
\end{equation}
where the two-dimensional alternating tensor has
$\epsilon^{12}=1$, and we have introduced
\be
\label{bracket-mom}
(1\: 2) \equiv (\p_1\:\p_2) := \hat 1 \tilde 2 - \hat 2 \tilde 1.
\ee
We can similarly express the contragredient invariant
$[\lambda_1\: \lambda_2] :=
\epsilon^{\dot\alpha\dot\beta}{\tilde\lambda}_{1\dot\alpha}{\tilde\lambda}_{2\dot\beta}$
in terms of the complex conjugate $2 {\{1\:2\}}/{\sqrt{\hat 1
\hat2}}$, where
\be
\label{barcket-mom}
\{1\:2\} := (1\:2)^* = \hat 1 \bar 2 - \hat 2 \bar 1.
\ee
Choice \eq{eq:spinor-choice} is not suitable for
$\mu_{\alpha\dot\alpha} = \nu_\alpha {\tilde\nu}_{\dot\alpha}$,
since the only non-zero covariant component is $\check\mu=1$. Thus
from \eq{quaternion} we take instead
\[
\nu=\tilde\nu =
\begin{pmatrix}\sqrt{2}\\0\end{pmatrix}.
\]
The standard polarisations for a massless on-shell vector boson of
momentum $p$,
\[
E_+ = {\nu\,\tilde\lambda\over\langle\nu\:\lambda\rangle}
\quad\text{and}\quad
E_- =
{\lambda\,\tilde\nu\over[\lambda\:\nu]},
\]
then have non-zero components
\be
\label{polarisations}
E_+ = -\frac 12, \quad \bar E_- = \frac 12.
\ee
There are also ${\check E}_+=-\bar p/2\hat p$ and ${\check E}_-=
p/2\hat p$, although these time-like components will not be
needed. (The remaining components vanish.)

%
%
Importantly, \eq{eq:spinor-choice} contains no reference to
$\check p$ and makes sense even for non-null momenta. In this
case, the spinors factor the bispinor associated to the null momentum $p+a\mu$,
where $a=  p\bar p/\hat p-\check p$. This definition is
equivalent to the CSW prescription for taking the spinors
off-shell \cite{CSW,Paul}, providing the CSW spinor is identified
(projectively) with $\nu$. Indeed, the CSW prescription is to
introduce a fixed spinor $\eta$ and take the off-shell momentum
spinor to be proportional to
\[
\epsilon^{\dot\alpha\dot
\beta}p_{\alpha\dot\alpha}{\tilde\eta}_{\dot\beta} =
\lambda_\alpha [\lambda\:\eta] - a\nu_\alpha [\nu\:\eta],
\]
so the two definitions coincide when $\eta\propto\nu$. We thus
arrive at the  satisfying conclusion that the arbitrary null
vector $\eta_\alpha{\tilde\eta}_{\dot\alpha}$ is just
$\mu_{\alpha\dot\alpha}$, the vector defining light-cone time and
the quantization surface.

\subsection{The transformation}

We take the Yang-Mills action written as \cite{Paul}
\be
S={1\over 2 g^2}\int\! dt\,dx^1dx^2dx^3\ {\rm tr}\, F^{\lambda
\rho}F_{\lambda \rho},
\ee
where
\be
F_{\lambda \rho}=[{D}_\lambda,\,{D}_\rho], \quad {
D}_\mu=\partial_\mu+A_\mu,\quad A_\mu=A^a_\mu T^a,
\ee
and the generators of the internal group have been taken as
anti-Hermitian:
\be
\label{generators}
[T^a,T^b]=f^{abc}T^c, \quad {\rm tr}\:(T^aT^b)=-{1\over 2}\,
\delta^{ab}.
\ee
We choose light-cone gauge $\hat A=0$, discarding the
non-interacting Fadeev-Popov ghosts, and integrate out the
longitudinal field $\check A$ (which appears quadratically and is
not dynamical, in the sense that the Lagrangian has no terms
$\check\partial\check A$).
%
%

The resulting action takes the form
\be
\label{action}
S={4\over g^2}\int\!\!dx^0 L,
\ee
where $L$ is the light-cone Lagrangian defined as an integral
$d^3{\bf x}=dx^{\bar 0}\,dz\,d{\bar z}$ over surfaces $\Sigma$ of
constant $x^0$. From \eq{polarisations}, $ A$ ($\bar A$) has only
positive (negative) helicity on-shell states. Labelling the 
parts by the participating helicities
$L=L^{-+}+L^{++-}+L^{--+}+L^{'--++}$, where
\bea
\label{ke}
L^{-+}[A] &=& \phantom{-}{\rm tr}\!\int_\Sigma\!\! d^3{\bf x}\,\,
{\bar A}\,(\check\partial\hat\partial-
\partial\bar\partial)\,A\\
\label{lclose}
{L}^{++-}[A]&=&-{\rm tr}\!\int_\Sigma\!\! d^3{\bf x}\,\,
({\bar\partial}{\hat\partial}^{-1} {  A})\,
[{  A},\,{\hat\partial} {\bar A}]\\
\label{lc3}
{L}^{--+}[A]&=&-{\rm tr}\!\int_\Sigma\!\! d^3{\bf x}\,\, [{\bar
A},\,{\hat\partial} {  A}]\,
({  \partial}{\hat\partial}^{-1} {\bar A})\\
\label{lc4}
{L}^{'--++}[A]&=&-{\rm tr}\!\int_\Sigma\!\! d^3{\bf x}\,\, [{\bar
A},\,{\hat\partial} { A}]\,{\hat\partial}^{-2}\, [{
A},\,{\hat\partial} {\bar A}].
\eea

Now we define a canonical change of variables from $A$ to $B$ to
absorb the unwanted term $L^{++-}$ into the kinetic term:
\be
\label{canondef}
L^{-+}[A]+{L}^{++-}[A]=L^{-+}[B].
\ee
This means that the transformation is performed on the
quantization surface $\Sigma$ with all fields having the same time
dependence $x^0$ (which henceforth we suppress). It induces the
following transformation for the canonical momenta:
\be
\label{canontrans}
{\hat\partial}{\bar A}^a({\bf y}) = \int_\Sigma\!\! d^3{\bf x}\,\,
{\delta B^b({\bf x})\over\delta A^a({\bf y})}\,
{\hat\partial}{\bar B}^b({\bf x}).
\ee
Substituting this back into \eq{canondef} yields the defining
relation between $A$ and $B$:
\be
\label{canonreln}
\int_\Sigma\! d^3\vec y \: \left[D,
\frac{\bar\partial}{\hat\partial} \:
    A \right]^a\!\!\!(\vec y) \:
    \frac{\delta B^b(\vec x)}{\delta A^a(\vec y)}
    = \frac{\partial \bar \partial}{\hat \partial}\, B^b(\vec
    x).
\ee
It follows that $A$ is a power series in $B$, of the form
$A=B+O(B^2)$ (at least for general momenta: see further comments
in \sec{explicit}) and thus from \eq{canontrans}, $\bar A$ is a
power series in $B$ each term containing also a single $\bar B$,
of the form $\bar A = \bar B + O(\bar BB)$.

Thus it follows from the equivalence theorem that we can equally
well use $B$ ($\bar B$) as the positive (negative) helicity field
in place of $A$ ($\bar A$) \cite{books,Paul}. The canonical nature
of the transformations ensures that the change of variables in the
functional integral has unit jacobian \cite{Paul}, whilst
substituting the transformations into $L$ results, from
\eq{canondef} and \eqs{lc3}{lc4}, in a Lagrangian with an infinite
number of interactions each containing just two $\bar B$ fields
and an increasing number of $B$ fields:
\be
\label{lb}
L = L^{-+}[B]+L^{--+}[B]+L^{--++}[B]+L^{--+++}[B]+\cdots.
\ee
This has precisely the structure required to be identified with
the MHV-rules Lagrangian, in the sense that its tree level
perturbation theory generates CSW rules with the Feynman rules
following from these vertices. It follows immediately that these
vertices when taken on-shell must be proportional to the
corresponding MHV amplitude since only one vertex is used to
construct the tree-level amplitude with the right helicity
assignment of two negative and any number positive helicities.
(Using two or more vertices results in ``N$^n$MHV'' amplitudes
with more than two negative helicities.)

It remains to show that when off-shell, these vertices must give
the CSW continuations of the MHV amplitudes. Mansfield argues that
this would follow from the fact that the vertices contain no
explicit $x^0$ dependence (or $\check\partial$) if one can show
that the vertices are also holomorphic in the sense that they
contain no $\bar\partial$ derivatives. We will see that this is
true at least for general momenta. Mansfield argues that this
holomorphy follows from considering the homogeneous
transformations:
  \be
\label{homog}
\delta A = [A,\theta] \ins11{and} \delta {\bar A} = [{\bar
A},\theta],
  \ee
where $\theta$ is a function of $\b z$ only. He uses the fact that
this corresponds to the shift $\delta \b\partial A =
[A,\b\partial\theta]$ when acting on \eq{canondef}, the equation
defining the expansion, and does not leave it invariant, whereas
\eq{homog} does leave invariant the sum of the two terms
\eqs{lc3}{lc4} which generates all the vertices in \eq{lb}. From
this viewpoint however, it is a surprise to find that the
coefficients of the series for $A$ ($\bar A$) themselves are
already holomorphic. Furthermore, we will show that they take a
very simple form.

\subsection{The precise correspondence}
\label{precise}

Clearly, it would be very welcome to investigate explicitly the
transformation \eqs{canontrans}{canonreln} and its effect on the
vertices. In order to do this, we write the general $n$-point term
in \eq{lb} in 3-momentum space as
\be
\label{vertex}
{1\over2}\sum_{s=2}^n \int_{12\cdots n}\!\!\!\!\!\!V^s_{12\cdots
n}\, \tr[{\bar B}_{\bar 1}B_{\bar 2}\cdots {\bar B}_{\bar s}
\cdots B_{\bar n}],
\ee
where the bar on the indices indicates that the 3-momentum dependence is
$B_{\bar k}\equiv B(-{\bf p}_k)$ (we continue to suppress their common $x^0$
dependence, which is not Fourier transformed), the missing fields in the trace
being $B_{\bar k}$s. The components of ${\bf p}$ are expressed, as always, as
$(\hat p,p,\bar p)$. The integral shorthand means, here and later,
\[
\int_{12\cdots n} \equiv\ \prod_{k=1}^n\int {d\hat p_k\:dp_k\:d\bar
p_k\over(2\pi)^3},
\]
and the vertex is expressed as
\be
\label{momfactor}
V^s_{12\cdots n}\ =\ (2\pi)^3\delta^3(\p_1+\p_2+\cdots+\p_n)\,
V^s(\p_1,\p_2,\cdots,\p_n).
\ee
Throughout the paper, expansion coefficients, vertices and
amplitudes carry these momentum conserving delta-functions; they
will be factored off in this manner and thus not written
explicitly. We often simply write $V^s(12\cdots n)\equiv
V^s(\p_1,\p_2,\cdots,\p_n)$. It should be borne in mind that from
\eq{momfactor} such coefficients are only defined when their
momentum arguments sum to zero.

Note that, by using \eq{generators}, we can always express the group theory
factors as traces of products of the $B$s and $\bar B$s valued in the Lie
algebra, as in \eq{vertex}. This form leads to the required colour-ordered
Feynman rules. We will express group theory factors throughout by absorbing them
in the fields, including for the expansion coefficients of \sec{Expansion}.

By the cyclicity of the trace, we can always arrange for the first field in
\eq{vertex} to be a $\bar B$. Since we have a choice of two $\bar B$s, we could
have restricted the sum in \eq{vertex} to $r\le\lfloor n/2+1 \rfloor$, however
by writing it as a full sum over $r$ and dividing the result by two, we get the
same thing except that the $\lfloor n/2+1 \rfloor^{\rm th}$ vertex is
accompanied by a factor $\half$ when $n$ is even, consistent with the fact that
it alone has a $Z_2$ symmetry under exchange of the $\bar B$s.

We would like to compare $V^s_{1\cdots n}$ to the MHV amplitude
\cite{ptbg}:
\be
\label{MHV}
A_n = g^{n-2}{\langle r\:s\rangle^4\over\langle 1\:2\rangle\langle
2\:3\rangle\cdots\langle n-1,n\rangle\langle n\:1\rangle},
\ee
$r$ and $s$ being the negative helicity legs, itself a component
of the full tree-level amplitude:
\be
\label{fullMHV}
\sum_{\sigma} {\rm tr}\, (T^{a_{\sigma(1)}}\cdots
T^{a_{\sigma(n)}})\: (2\pi)^4{ i}\:\delta^4(
p^1+\cdots+p^n)\,A_n^\sigma,
\ee
the sum being over distinct cyclic orderings $\sigma$. However
this is written with different conventions.

There are many ways to perform the translation. Perhaps the
following is simplest. The normalization for the generators in
\eqs{MHV}{fullMHV} is such that $\tr\,(T^a T^b)= \delta^{ab}$ and
$[T^a,T^b] ={ i}\sqrt{2}f^{abc}T^c$ \cite{Dixon:1996wi}. Comparing
with \eq{generators} we see we need to replace $T^a \mapsto -{
i}T^a/\sqrt{2}$. To form the momentum space Feynman rule from
\eq{vertex} and \eq{momfactor}, we Fourier transform the $x^0$
dependence also, obtaining a four-dimensional delta-function as in
\eq{fullMHV}, however in our case it is defined via \eq{coord} to
be
\[
(2\pi)^4\delta^4(p) = \int\! dx^0dx^{\bar 0}dz\,d{\bar z}\ \, {\rm
e}^{\displaystyle \,{ i}(\check p\,x^0+\hat p\,x^{\bar 0}+p\,z+\bar
p\,\bar z)}.
\]
As can be seen by computing the jacobian or $\sqrt{-g}$, this is
four times the one in \eq{fullMHV}. Noting that $p^\mu p_\mu =
4(\check p\hat p - p\bar p)$, we see that \eq{action} and \eq{ke}
yield the propagator $2ig^2\delta^{ab}/p^2$. To bring this to
canonical normalization requires absorbing $2g^2$ by $B\mapsto
Bg\sqrt{2}$ (similarly $\bar B$). Combining all these with the
prefactor from \eq{action}, we see that the $r=1$ MHV amplitude
should be given by
\[
{16\over g^2}\, (-{ i}g)^n\, V^s(12\cdots n)\, (E^+)^{n-2}\,(\bar
E^-)^2.
\]
Of course the factor $\half$ for the even-$n$ $\lfloor n/2+1
\rfloor^{\rm th}$ vertex in \eq{vertex} is cancelled here by the
two ways to form this amplitude. Finally, from \eq{polarisations},
\eq{MHV} and \eq{eq:anglebracket-mom} we thus expect to find
\bea
V^s(12\cdots n) &=& (2{ i})^{n-4}\,\ {\langle
1\:s\rangle^4\over\langle 1\:2\rangle\langle
2\:3\rangle\cdots\langle n-1,n\rangle\langle n\:1\rangle}\nonumber\\
&=& { i}^n
{\hat2\cdots\hat{n}\,(1\:s)^4\over\hat1\hat{s}^2(1\:2)(2\:3)\cdots(n-1,n)(n\:1)}.
\label{V}
\eea

\section{Explicit Expansion}
\label{Expansion}
\label{explicit}

In this section we derive recursion relations for the expansions
satisfying \eq{canonreln} and \eq{canontrans} in the case that $A$
and $B$ have support only on general momenta. We then solve these
and use the results to confirm \eq{V} for the three point, the two
four-point and the two five-point vertices.

\subsection{$A$ expansion coefficients to all orders}

Rearranging \eq{canonreln} and transforming to 3-momentum space
yields
\be
\label{momcanonreln}
\omega_1 A_1 -{ i}\int_{23}\zeta_3[A_2,A_3]\,
(2\pi)^3\delta^3(\p_1-\p_2-\p_3) = \int_\p\omega_\p B^b_\p{\delta
A_1\over\delta B^b_\p},
\ee
using the notation set up in sub\sec{precise}, and introducing
$\zeta_{\vec{p}}\equiv\zeta(\p) ={\bar p}/{\hat p}$ and
$\omega_{\vec{p}} = {p \bar p}/{\hat p}$.

From this one would be tempted to conclude that at lowest order
$A_\p = B_\p$ and thus in general, absorbing the group theory
generators into the fields, $A$ has an expansion in $B$ of the
form
\be
\label{Aexp}
A_1 = \sum_{n=2}^\infty\int_{2\cdots n}\!\!\!\!\Upsilon_{12\cdots
n}\,B_{\bar2}\cdots B_{\bar n},
\ee
where $\Upsilon(\p,-\p)=1$. However we have to divide through by
$\omega_\p$ to conclude that $A_\p = B_\p$ to lowest order, and
since $\omega_\p$ vanishes when $p=0$ or $\b p=0$, more general
solutions exist where $A_\p$ has a piece not containing $B_\p$ but
with delta-function support, \viz $\delta(\b p)$ and/or
$\delta(p)$.

It is tempting to ignore these terms, however we cannot do so and
also implement some expected residual gauge invariances: the light
cone action \eq{action} is arrived at by fixing the gauge
$\mu\cdot A \propto {\hat A} = 0$. This leaves gauge
transformations $\delta A_\mu = [D_\mu,\theta]$ unfixed providing
$\theta$ does not depend on ${\hat z}$. If $\theta$ depends on
$x^0$, we can expect the form of the gauge transformation to be
modified as a result of $\check A$ being integrated out. On the
other hand, it is easy to check that \eq{action} is invariant
under holomorphic gauge transformations $\theta(z)$, providing we
interpret\footnote{We note that this is consistent with the
Mandelstam-Leibbrandt prescription \cite{Mandle}.}
$\b\partial{\hat\partial}^{-1}\theta:=
{\hat\partial}^{-1}(\b\partial\theta)=0$. Similarly (by symmetry)
we also have antiholomorphic gauge invariance generated by
$\theta(\b z)$.

Now, the left hand side of \eq{canondef} is actually invariant
under holomorphic gauge transformations which indicates that $B$
and $\b B$ must also be invariant. Indeed this ensures that the
defining transformations \eqs{canontrans}{canonreln} transform
covariantly. However, \eq{Aexp} only transforms covariantly if we
allow for an extra term independent of $B$ and proportional to
$\delta(\b p_1)$ to absorb the gauge transformation. This
indicates one should really interpret \eq{Aexp} as amounting to
further gauge fixing. For the present we will simply declare that
\eq{Aexp} is to be applied only for generic momenta in particular
such that $p_1$ and $\b p_1$ are non-zero.

Substituting \eq{Aexp} in \eq{momcanonreln}, comparing
coefficients and stripping off momentum conserving delta functions
as in \eq{momfactor}, yields the recurrence relation:
\be
\label{UpsilonR}
\Upsilon(1\cdots n) = {{ i}\over\omega_1
+\cdots+\omega_n}\sum^{n-1}_{j=2}\left(\zeta_{j+1,n}-\zeta_{2,j}\right)
\Upsilon(- ,2,\cdots,j)\Upsilon(- ,j+1,\cdots,n),
\ee
where the arguments labelled ``$-$'' are minus the sum of the
remaining arguments (as follows from momentum conservation) and we
have defined $\zeta_{j,k}=\zeta(P_{j,k})$ with
$P_{j,k}=\sum_{i=j}^k \p_i$ (similarly for $\omega_{j,k}$ below).
The next two coefficients are thus
\bea
\label{Upsilon3}
\Upsilon(123) &=&
   { i} \frac{\zeta_3 - \zeta_2}
         {\omega_{1} + \omega_2 + \omega_3}, \\
\label{Upsilon4}
\Upsilon(1234) &=& \frac 1 {\omega_{1} + \omega_2 + \omega_3 +
\omega_4} \left[
    \frac{(\zeta_4 - \zeta_3) (\zeta_{3,4} - \zeta_2)}
         {\omega_{3,4} - \omega_3 - \omega_4}
  + \frac{(\zeta_{4} - \zeta_{2,3})(\zeta_3 - \zeta_2) }
         {\omega_{2,3} - \omega_2 - \omega_3}
\right]\!.
\eea
Although apparently not holomorphic, once they are expressed in
terms of independent momenta and simplified, all $\b p_k$
dependence drops out, resulting in very compact expressions:
\begin{eqnarray*}
\Upsilon(123)   &=& -{ i} {{\hat1}\over(2\: 3)}, \\
\Upsilon(1234)  &=& {\hat1\hat3\over(2\:3)(3\:4)},\\
\Upsilon(12345) &=& { i} {\hat1\hat3\hat4\over(2\:3)(3\:4)(4\:5)}.
\end{eqnarray*}
We have displayed the four-point coefficient which we also checked
explicitly. Now we prove by induction that the general coefficient
takes the form
\be
\label{Upsilonn}
\Upsilon(1\cdots n) = i^n {\hat1\hat3\hat4\cdots
\widehat{n-1}\over(2\:3)(3\:4)\cdots(n\!-\!1, n)},
\ee
where $n\ge4$. It is sufficient to show that substituting
\eq{Upsilonn} into the right hand side of \eq{UpsilonR} yields the
left hand side. Substituting \eq{Upsilonn} into the right hand
side yields
\[
-{\Upsilon(1\cdots n)\over\hat1(\omega_1+\cdots+\omega_n)}
\sum^{n-1}_{j=2} {(j, j\!+\!1)\over\hat{
j\,}\widehat{j\!+\!1}}\,\{P_{2,j}\ P_{j+1,n}\},
\]
after due care with the ends of the sum. Expanding the $(j,
j\!+\!1)$ term and relabelling so that both halves are indexed by
$j$ converts the sum to
\[
\sum^n_{j=2}{\tilde j\over\hat j}\,\left(\{P_{2,j-1}\ P_{j,n}\} -
\{P_{2,j}\ P_{j+1,n}\}\right),
\]
where contributions at the end of the sum are correctly
incorporated by defining $P_{j,k}=0$ when $j>k$. Writing $P_{2,j}
= P_{2,j-1}+\p_j$ and $P_{j+1,n}=-P_{2,j-1}-\p_1-\p_j$, and using
the antisymmetry of \eq{barcket-mom}, the sum collapses to
$-\hat1(\omega_1+\cdots+\omega_n)$, proving the assertion
\eq{Upsilonn}.

\subsection{$\b A$ expansion coefficients to all orders}

Differentiating \eq{Aexp} with respect to $B$ and substituting the
inverse into \eq{canontrans} yields an expansion for $\bar A$ of
the form
\be
\label{Abexp}
{\hat1}\A_\b1 = \sum_{m=2}^\infty\,\sum_{s=2}^m\int_{2\cdots
m}\!\!\!\!{\hat s}\,\,\Xi^{s-1}_{\b12\cdots m}\,B_\b2\cdots\B_\b
s\cdots B_\b m,
\ee
where the superscript on $\Xi$ labels the relative position (not
momentum) of the $\B$ field and the missing fields are $B_\b ks$.
We use the invariant
\[
{\rm tr} \int_{\Sigma}\!\! d^3\vec{x} \ \check\partial
A\hat{\partial} \bar A = {\rm tr} \int_{\Sigma}\!\! d^3\vec{x} \
\check\partial B\hat{\partial} \bar B
\]
to extract a recurrence relation. Recalling that all fields have
the same $x^0$ dependence, we have from \eq{Aexp},
\[
\check\partial A_1 = \sum_{n=2}^\infty\sum_{r=2}^n\int_{2\cdots
n}\!\!\!\!\Upsilon_{12\cdots n}\,B_{\bar2}\cdots \check\partial
B_\b r\cdots B_{\bar n}.
\]
Substituting this and \eq{Abexp} into the above, using cyclicity
of the trace and several careful relabellings we find:
\begin{multline}
\label{XiR}
\Xi^l(1\cdots n) = -\!
\sum_{r=2}^{n+1-l}\!\sum_{m=\max(r,3)}^{r+l-1}
    \Upsilon(-,n\!-\!r\!+\!3,\cdots,m\!-\!r\!+\!1) \times \\
    \Xi^{l+r-m}(-,m\!-\!r\!+\!2,\cdots,n\!-\!r\!+\!2),
\end{multline}
where $l=1,\cdots,n-1$, the momentum indices on the right hand
side must be interpreted cyclically, \ie mod $n$, and
$\Xi^1(\p,-\p)=1$. [Note that $m$ is the number of arguments in
$\Upsilon$. The inner sum should be interpreted as zero when $r=2$
and $l=1$, alternatively the lower limit in $r$ can be given as
$\max(2,4-l)$.] From this, or directly, one can readily compute
the first few coefficients:
\begin{align}
\Xi^1(123) &= -\Upsilon(231), \notag \\
\Xi^2(123) &= -\Upsilon(312), \notag \\ 
\Xi^1(1234)&=
-\Upsilon(2\!+\!3,4,1)\,\Xi^1(1\!+\!4,2,3)-\Upsilon(2341),
    \notag \\
\Xi^2(1234) &= -\Upsilon(3\!+\!4,1,2)\,\Xi^1(1\!+\!2,3,4)
                -\Upsilon(2\!+\!3,4,1)\,\Xi^2(1\!+\!4,2,3)-\Upsilon(3412),
    \notag \\
\Xi^3(1234) &=
-\Upsilon(3\!+\!4,1,2)\,\Xi^2(1\!+\!2,3,4)-\Upsilon(4123).
\notag 
\end{align}
Clearly, since the $\Upsilon$ coefficients are already
holomorphic, the $\Xi$s will turn out to be also. In fact they
take a very simple form when expressed in terms of $\Upsilon$:
\be
\label{Xijn}
\Xi^{s-1}(1\cdots n) = - {\hat s \over\hat1} \Upsilon(1\cdots n),
\ee
($s=2,\cdots,n$ and $n\ge2$).

Let us now prove this assertion. Again, it is sufficient to show
that substituting \eq{Xijn} into the right hand side of \eq{XiR}
yields its left hand side. Substituting \eq{Xijn} into the right
hand side and using \eq{Upsilonn}, we find we can extract a factor
of $\Xi^l(1\cdots n)$ and thus learn that proving \eq{Xijn} is
equivalent to proving that
\[
\sum_{r=2}^{n+1-l}\!\sum_{m=\max(r,3)}^{r+l-1} {(m\mi r\pl1,\ m\mi
r\pl2) (n\mi r\pl2,\ n\mi r\pl3) {\hat P}_{n-r+3, m-r+1}\over
\widehat{m\mi r\pl1}\ \widehat{m\mi r\pl2}\ \widehat{n\mi r\pl2}\
\widehat{n\mi r\pl3}}
\]
equals
\be
\label{horatio}
-{(1\:2)(n\:1)\over\hat1\hat2\hat n}
\ee
In a similar way to the proof of \eq{Upsilonn}, we now expand the
factor $(m\mi r\pl1,\ m\mi r\pl2)$ and relabel so that both halves
are collected. (However since here momentum labels are treated mod
$n$, ${\hat P}_{j,k}$ means summing from $j$ increasing to $k$,
going through $1$ when $k<j$.) This allows us to perform the inner
sum with the result displayed in braces:
\be
\label{outer}
\sum_{r=2}^{n+1-l} {(n\mi r\pl2,\ n\mi
r\pl3)\over\widehat{n\mi r\pl2}\ \widehat{n\mi r\pl3}}\left\{
-{\tilde P}_{q+2,l}+{\widetilde{l+1}\over\widehat{l+1}}\,{\hat
P}_{n-r+3,l}-{\widetilde{q+1}\over\widehat{q+1}}\,{\hat
P}_{n-r+3,\, q+1}\right\}.
\ee
Here $q=1$ if $r=2$ otherwise $q=0$. We note that the term in
curly brackets vanishes when $r=2$ and $l=1$, as required, \cf
below \eq{Xijn}. At this stage it proves useful to consider the
case $l=n-1$ separately. Substituting this into \eq{outer} gives
\[
{(n\:1)\over\hat1\hat n}\left\{{\tilde n}+\tilde1+\tilde2\
-{\tilde n} -{\tilde2\over\hat2}{\hat P}_{1,2}\right\},
\]
where we have used momentum conservation on the first two terms in
curly brackets. It is immediate to see that this gives
\eq{horatio} as required. For $l<n-1$, we expand $(n\mi r\pl2,\
n\mi r\pl3)$ and collect both halves. This gives
\begin{multline*}
{\tilde
P}_{1,l+1}\left({\tilde1\over\hat1}-{\widetilde{l+1}\over\widehat{l+1}}\right)
+{\tilde n\over\hat
n}\left({\hat1\tilde2\over\hat2}-\tilde1\right)+{\tilde1\over\hat1}\left(-{\tilde
P}_{3,l}+{\widetilde{l+1}\over\widehat{l+1}}{\hat
P}_{1,l}-\tilde2-{\hat1\tilde2\over\hat2}\right)\\
+{\widetilde{l+1}\over\widehat{l+1}}\left({\tilde
P}_{2,l}+\widetilde{l+1}+{\tilde1\over\hat1}{\hat
P}_{l+2,1}\right).
\end{multline*}
Using momentum conservation and cancelling terms, this expression
collapses to \eq{horatio}, and thus \eq{Xijn} is proved.

Recall that to obtain the vertices \eqs{vertex}{momfactor} in
\eq{lb}, we substitute the series \eq{Aexp} and \eq{Abexp}, which
we have seen are holomorphic, into \eq{lc3} and \eq{lc4}, which
are also holomorphic. Thus we have proven that the vertices in
\eq{lb} are holomorphic. It follows by Mansfield's arguments
\cite{Paul} that these vertices when off-shell give the CSW
continuations of the MHV amplitudes, \ie are the ones in \eq{V}.
Nevertheless it is instructive to verify that this really does
work out in practice.

\subsection{Three-point vertex}

Since $A$ ($\bar A$) is, to lowest order, linear in $B$ ($\bar
B$), the three-point vertex is simply the light-cone gauge vertex
\eq{lc3}. Transforming to 3-momentum space and casting in the form
\eq{vertex}, we have
\be
\label{V3}
V^2(123) = { i} \: \frac{\hat 3}{\hat 1 \hat 2} (2\: 1).
\ee
On the other hand, from \eq{V} we expect
\[
-{ i}{\hat3\over\hat1\hat2}{(1\:2)^3\over(2\:3)(3\:1)}.
\]
Substituting $\p_3=-\p_1-\p_2$ in the denominator, we readily see
that these equations are the same and thus simply verify that the
light-cone gauge three-point vertex satisfies the general formula
\eq{V} as expected:
\[
V^2(123) = {1\over2{ i}} \frac{\langle 1\:2 \rangle^3}{\langle 2\:3
\rangle\langle 3\:1 \rangle}.
\]

\subsection{Four-point vertices}

The four-point vertices receive contributions from \eq{lc3} and
the first non-trivial terms in \eqs{Aexp}{Abexp}, and also
directly from \eq{lc4}, and thus are more exacting tests. We write
\eq{lc4} in the same form as \eq{vertex}:
\[
L^{'--++} = \int_{1234}\left\{\ W^2_{1234}\,
\tr[\A_\b1\A_\b2A_\b3A_\b4] +
{1\over2}\,W^3_{1234}\,\tr[\A_\b1A_\b2\A_\b3A_\b4]\ \right\},
\]
where
\[
W^2(1234) =
-{\hat1\hat3+\hat2\hat4\over(\hat1+\hat4)^2}\ins11{and} W^3(1234)
= {\hat1\hat4+\hat2\hat3\over(\hat1+\hat2)^2}
+{\hat1\hat2+\hat3\hat4\over(\hat1+\hat4)^2}
\]
(after symmetrization). Therefore,
\be
\label{V21234}
V^2(1234) = {\hat1\over\hat5}V^2(523)\,\Xi^2(\b541) +
{\hat2\over\hat5}V^2(154)\,\Xi^1(\b523) +V^2(125)\Upsilon(\b534) +
W^2(1234),
\ee
where the momentum $\p_5$ is determined by conservation in each
term (thus \eg in the first term $\p_5=\p_1+\p_4$). To compare
this formula to the expected result \eq{V} we map both to unique
functions of independent momenta. For example, we substitute for
the last momentum: $\p_4 = -\p_1-\p_2-\p_3$. It is then
straightforward using computer algebra to show that this coincides
with the right hand side of \eq{V}, and thus:
\[
V^2(1234) =
{\langle1\:2\rangle^3\over\langle2\:3\rangle\langle3\:4\rangle\langle4\:1\rangle}.
\]
For example, simplifying by partial fractions, both \eq{V} and
\eq{V21234} give:
\[
\frac{(\hat 1 + \hat 2)^2 (\hat 1 \tilde 2 - \hat 2 \tilde 1)}
    {\hat 1 \hat 2 \: [(\hat 1 + \hat 2) \tilde 3 - \hat 3 \tilde 1
                                                  - \hat 3 \tilde
                                                  2]}
-\frac{\hat 2 (\hat 1 + \hat 2 + \hat 3)(\hat 1 \tilde 2 - \hat 2
\tilde 1)}
    {\hat 1 (\hat 2 + \hat 3) (\hat 2 \tilde 3 - \hat 3 \tilde 2)}
-\frac{\hat 1 \hat 3 (\hat 1 \tilde 2 - \hat 2 \tilde 1)}
    {\hat 2 (\hat 2 + \hat 3) (\hat 1 \tilde 2 + \hat 1 \tilde 3
         - \hat 2 \tilde 1 - \hat 3 \tilde 1) }.
\]
Similarly, after symmetrization,
\begin{eqnarray*}
  V^3(1234) = && \hatfrac15 \: V^2(352)\, \Xi^2(\bar 541) +
    \hatfrac 35 \: V^2(512)\, \Xi^1(\bar 534) \\
   &+& \hatfrac 35 \: V^2(154)\, \Xi^2(\bar 523) +
    \hatfrac 15 \: V^2(534)\, \Xi^1(\bar 512) + W^3(1234),
\end{eqnarray*}
and it is straightforward to confirm as above that this agrees
with \eq{V}:
\[
V^3(1234) \ =\ {\hat2\hat4\over\hat1\hat3}\,\,
{(1\:3)^4\over(1\:2)(2\:3)(3\:4)(4\:1)}.
\]

\subsection{Five-point vertices}

The five-point vertices leave no doubt that the off-shell MHV
vertices are produced: they involve substituting up to the first
three terms in the expansions \eqs{Aexp}{Abexp} into both original
vertices \eqs{lc3}{lc4}. We find
\begin{eqnarray*}
V^2(12345) &=&
    \hatfrac 36 \: V^2(612) \,\Xi^1(\bar 6345)
  + \hatfrac 17 \: V^2(726) \Upsilon(\bar 634) \,\Xi^2(\bar 751)
  + V^2(126) \Upsilon(\bar 6345)\\
  &+& \hatfrac 26 \hatfrac 17 \: V^2(764) \,\Xi^1(\bar 623) \,\Xi^2(\bar 751)
  + \hatfrac 26 \: V^2(167) \,\Xi^1(\bar 623) \Upsilon(\bar 745)\\
  &+& \hatfrac 26 \: V^2(165) \,\Xi^1(\bar 6234)
+ W^2(1236) \Upsilon(\bar 645)
  + W^2(1265) \Upsilon(\bar 634)\\
  &+& \hatfrac 26 \: W^2(1645) \,\Xi^1(\bar 623)
  + \hatfrac 16 \: W^2(6234) \,\Xi^2(\bar 651),\\
V^3(12345) &=&
    \hatfrac 36 \: V^2(612) \,\Xi^1(\bar 6345)
  + \hatfrac 16 \: V^2(634) \,\Xi^2(\bar 6512)
  + \hatfrac 16 \: V^2(637) \,\Xi^1(\bar 612) \Upsilon(\bar 745)\\
  &+& \hatfrac 16 \hatfrac 37 \: V^2(675) \,\Xi^1(\bar 612) \,\Xi^1(\bar 734)
  + \hatfrac 17 \hatfrac 36 \: V^2(672) \,\Xi^1(\bar 634) \,\Xi^2(\bar
  751)\\
  &+& \hatfrac 16 \hatfrac 37 \: V^2(674) \,\Xi^2(\bar 651) \,\Xi^2(\bar 723)
  + \hatfrac 36 \: V^2(167) \,\Xi^2(\bar 623) \Upsilon(\bar 745)\\
  &+& \hatfrac 36 \: V^2(165) \,\Xi^2(\bar 6234)
  + \hatfrac 16 \: V^2(362) \,\Xi^3(\bar 6451)
+ \hatfrac 36 \: W^2(1645) \,\Xi^2(\bar 623)\\
  &+& \hatfrac 16 \: W^2(6345) \,\Xi^1(\bar 612)
  + \hatfrac 36 \: W^3(1265) \,\Xi^2(\bar 634)
  + \hatfrac 16 \: W^3(3462) \,\Xi^2(\bar 651)\\
  &+& W^3(1236) \Upsilon(\bar 645)
\end{eqnarray*}
(where, like before, indices 6 and 7 label momenta that are
uniquely determined in terms of the first five by momentum
conservation). Again, eliminating $\p_5$ in favour of the first
four momenta and doing likewise for the corresponding right hand
sides in \eq{V}, we find the expressions agree, and thus confirm
that
\begin{alignat*}{2}
V^2(12345) &= 2{ i} \frac{\langle 1\:2 \rangle^3}
    {\langle 2\:3 \rangle \langle 3\:4 \rangle
     \langle 4\:5 \rangle \langle 5\:1 \rangle},
&\qquad&\text{and}\\
V^3(12345) &= 2{ i} \frac{\langle 1\:3
\rangle^4}
    {\langle 1\:2 \rangle \langle 2\:3 \rangle \langle 3\:4 \rangle
     \langle 4\:5 \rangle \langle 5\:1 \rangle}.
\end{alignat*}

\section{Equivalence Theorem Matching Factors}
\label{ET}

Even if a change of field variables from $A$ to $B$ turns the
light-cone gauge Lagrangian into an MHV-rules Lagrangian, this
does not mean necessarily that one-loop and higher contributions
are obtained purely by using the CSW rules. We have to remember
that wavefunction renormalization terms, which we have already
referred to as ET matching factors, are also generated
\cite{books,ET}.

Let us recall that an S-matrix element is obtained by computing
the amputated Green function (the LSZ procedure). Thus to the
action \eq{action}, we should add the source terms
\be
\label{sources}
\int\!\! d^4x\, J^a\b A^a + \b J^a A^a,
\ee
where, after amputating propagators $\langle A\: \b A\rangle(p)$,
by multiplying by $-i p^2$ and taking the on-shell limit $p^2\to
0$, we see that $J$ acts as a source for positive helicity $A$
legs in the amputated Green function (and likewise $\b J$ for
negative helicity $\b A$ legs).

Now, substituting the series \eq{Aexp} and \eq{Abexp} into
\eq{sources}, we see that at tree-level the process of multiplying
by $-ip^2$ and taking the on-shell limit, kills all terms but the
first, which survives because these generate cancelling poles via
the propagators $\langle B\: \b B\rangle(p)$. Therefore, we can
effectively write the source terms as
\[
\int\!\! d^4x\, J^a\b B^a + \b J^a B^a.
\]
This is the tree-level content of the Equivalence Theorem.

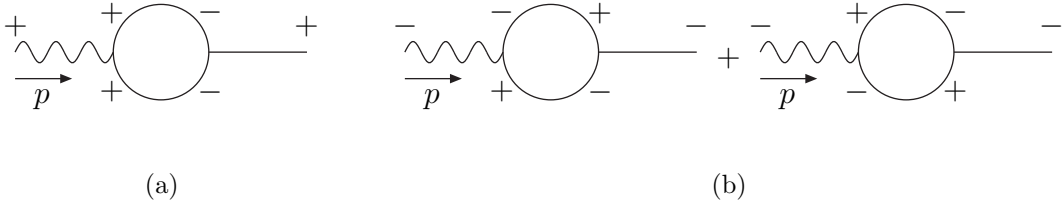
\begin{figure}[!h]
\centering
\subfigure[]{
  \(\begin{matrix}\begin{picture}(120,60)
    \SetOffset(60,30)

    \CArc(0,0)(18,0,360)
    \Photon(-55,0)(-18,0){4}{3}
    \Line(18,0)(55,0)

    \Text(-55,6)[bc]{$+$}
    \Text(+55,6)[bc]{$+$}
    \Text(-14,11)[br]{$+$}
    \Text(-14,-11)[tr]{$+$}
    \Text(14,11)[bl]{$-$}
    \Text(14,-11)[tl]{$-$}

    \LongArrow(-55,-10)(-35,-10)
    \Text(-45,-14)[tc]{$p$}
  \end{picture}\end{matrix}\)
  \label{oneloop-a}
}
\quad
\subfigure[]{
  \(\begin{matrix}\begin{picture}(120,60)
    \SetOffset(60,30)

    \CArc(0,0)(18,0,360)
    \Photon(-55,0)(-18,0){4}{3}
    \Line(18,0)(55,0)

    \Text(-55,6)[bc]{$-$}
    \Text(+55,6)[bc]{$-$}
    \Text(-14,11)[br]{$-$}
    \Text(-14,-11)[tr]{$+$}
    \Text(14,11)[bl]{$+$}
    \Text(14,-11)[tl]{$-$}

    \LongArrow(-55,-10)(-35,-10)
    \Text(-45,-14)[tc]{$p$}
  \end{picture}\end{matrix}
  +
  \begin{matrix}\begin{picture}(120,60)
    \SetOffset(60,30)

    \CArc(0,0)(18,0,360)
    \Photon(-55,0)(-18,0){4}{3}
    \Line(18,0)(55,0)

    \Text(-55,6)[bc]{$-$}
    \Text(+55,6)[bc]{$-$}
    \Text(-14,11)[br]{$+$}
    \Text(-14,-11)[tr]{$-$}
    \Text(14,11)[bl]{$-$}
    \Text(14,-11)[tl]{$+$}

    \LongArrow(-55,-10)(-35,-10)
    \Text(-45,-14)[tc]{$p$}
  \end{picture}\end{matrix}\)
  \label{oneloop-b}
}
\caption{Topology of the one-loop ET matching
factors, with \subref{oneloop-a} the contribution to negative helicity
$\b B$ legs; and \subref{oneloop-b} the contribution to positive helicity
$B$ legs. Wavy lines denote $A$ and $\b A$ fields, straight lines
$B$ and $\b B$.}
\label{oneloop}
\end{figure}

At one loop and higher, the only change to this conclusion is that
we can use the vertices in the expansions \eqs{Aexp}{Abexp} and
those of the Lagrangian, \viz \eq{vertex}, to form
self-energy-like diagrams, \cf \fig{oneloop}. Since these again
attach to propagators carrying the external momentum $p^\mu$, they
survive the process of amputation.

These ET matching factors are not explicitly Lorentz invariant, a
consequence of both the light-cone axial gauge and the change to
$B$ fields. However, if everything is defined correctly, Lorentz
invariance should be recovered in the process of forming S-matrix
elements. Therefore we should find that these matching factors
depend only on $p^2$, which is sent to zero. Since these
self-energy-like diagrams then depend on no scale at all, in
dimensional regularisation we must set them to zero.

We now confirm these conclusions at one loop. The terms in
\fig{oneloop} contribute by multiplying the legs of the
corresponding tree-level amplitude (computed using the CSW rules).
The diagrams in \fig{oneloop} contain both ultraviolet and
infrared divergences. We derived the vertices \eqs{Aexp}{Abexp}
and \eq{vertex} without using any regularisation. Strictly
speaking, we should therefore wait until a corresponding
Lagrangian is supplied incorporating sufficient regularisation.
However, if we proceed with the calculation, na\"\i vely applying
dimensional regularisation where needed, we can expect to get
right the most divergent pieces.

From the diagram \fig{oneloop-a}, using \eq{Aexp} and \eq{V3}, we obtain
\be
\label{diaga}
-{1\over2} g^2C_A\int\!\! {d^4q\over(2\pi)^4}\,
{\Upsilon(-p,p+q,-q) V^2(q,-p-q,p)\over q^2(p+q)^2} = -{ i}\hat
p\,g^2C_A\int\!\! {d^4q\over(2\pi)^4}\, {1\over q^2(p+q)^2\hat q}
\ee
where a factor $4g^2$ arises from the multiplier in \eq{action}, a
factor $1/4$ arises for converting $dq\,d\b q\,d\hat q\,d\check q$
to the $dq^tdq^1dq^2dq^3$ taken above, and $C_A/2$ arises from
evaluating the trace over the various products of generators,
$C_A$ being the adjoint Casimir [so $C_A=N$ for $SU(N)$]. The
right hand side follows after using partial fractions and shifts
in loop momentum. We can write it in Lorentz invariant fashion
using $\mu$, and furthermore map it to $D=4-2\epsilon$ dimensions:
\be
\label{kos}
-{ i}g^2C_A\,\mu\!\cdot\!p\int\!\! {d^Dq\over(2\pi)^D}\, {1\over
q^2(p+q)^2\mu\!\cdot\!q}.
\ee
In order to evaluate the integral we need to keep $p$ off-shell
and let $p^2\to0$ at the end. This is consistent with the way this
contribution arises through the LSZ procedure.

The integral has already been evaluated in ref. \cite{david} and
we reproduce the result in the appendix. However we can see that
it vanishes in the $p^2\to0$ limit without calculation. By Lorentz
invariance, the result can only depend on $p^2$ and
$\mu\!\cdot\!p$ ($\mu^2$ being zero). However since the expression
is independent of $r$ when scaled as $\mu\mapsto r\mu$, we see
that the result does not in fact depend on $\mu\!\cdot\!p$. We
thus confirm that the integral depends only on the scale $p^2\to0$
and thus must be set to zero, as required in dimensional
regularisation.\footnote{In more detail, by dimensions the
integral depends only on $p^2$ through the factor
$(-p^2)^{-\epsilon}$. The vanishing of the integral is then fully
justified if we keep $\epsilon<0$ as $p\to0$.}

Now we turn to the two diagrams of \fig{oneloop-b}. Using
\eq{Abexp} and \eq{V3}, we obtain
\bea
&&{1\over2}g^2C_A\int\!\! {d^4q\over(2\pi)^4}\, {1\over
q^2(p+q)^2} \left(1+{\hat q\over\hat
p}\right)\Upsilon(p+q,-q,-p)V^2(p,q,-p-q)\nonumber\\
&-&{1\over2}g^2C_A\int\!\! {d^4q\over(2\pi)^4}\, {1\over
q^2(p+q)^2}\,{\hat q\over\hat
p}\,\Upsilon(-q,-p,p+q)V^2(-p-q,p,q),\nonumber
\eea
which simplifies to
\be
\label{diagb}
-{ i}g^2C_A\int\!\! {d^4q\over(2\pi)^4}\, {1\over
q^2(p+q)^2}\,{(\hat p+\hat q)^3\over\hat q\hat p^2}.
\ee
Writing this in a Lorentz invariant way in $D$ dimensions gives
\be
\label{stuff}
-ig^2C_A\int\!\! {d^Dq\over(2\pi)^D}\, {1\over q^2(p+q)^2}\,
{(\mu\!\cdot\!p+\mu\!\cdot\!q)^3\over\mu\!\cdot\!q(\mu\!\cdot\!p)^2}\,.
\ee
This integral is also evaluated off shell in the appendix, however
again we see that this expression is independent of
$\mu\!\cdot\!p$ and thus depends only on the scale $p^2$, forcing
the integral to vanish on shell, in dimensional regularisation.

Finally we would like to note that in this case the contribution
contains the term
\[
{11\over6}{g^2C_A\over(4\pi)^2}{1\over\epsilon}
(-p^2)^{-\epsilon},
\]
arising from an ultraviolet divergence. It is tempting to try and
interpret this in the case where $p^2$ is small but non-zero,
since if we add the divergence to the $B$ tree-level matching
factor (which is just $Z_+=1$), it is precisely of the correct
form to be cancelled by renormalizing a factor of the bare
coupling constant $g_0$, such as appears in the tree-level
MHV-rules amplitude. The left-over $\ln(-p^2)$ would be expected
to cancel such a term in tree-level bremsstrahlung from off-shell
positive helicity gluons. However this cannot be the whole story.
In particular we expect other divergences in the one-loop
amplitudes obtained using MHV-rules.

\section{Conclusions}
\label{Conclusions}

We have seen that the direct series solution of the transformation
proposed in ref.~\cite{Paul} does indeed yield the vertices \eq{V}
necessary to reproduce the CSW rules at tree level.

A key step in making the expansions manageable was to recognize
that all the group theory factors could be absorbed into the
fields, allowing also the derivation of compact recurrence
relations \eqs{UpsilonR}{XiR}. We were able to solve these and
thus discovered very simple expressions \eqs{Upsilonn}{Xijn} for
the coefficients of the expansion of $A$ and $\b A$, in terms of
the ``MHV-frame'' $B$ field to all orders. We expect that these
explicit expressions will be useful for further developments in
the subject. Of particular note is that the expansion coefficients
are purely holomorphic
--- a property we would like to understand on a more profound
basis than we do at present. It follows that all the Lagrangian
vertices \eq{vertex} are also holomorphic. We can then use the
arguments as given in ref.~\cite{Paul} to prove that all these
vertices must correspond to the CSW off-shell extension of
MHV-rules vertices.

We would like to mention that the comparison between the explicitly derived
vertices that come out of the expansion and MHV vertices was very
straightforward to do algebraically once the spinors were converted into
momentum components using \eq{eq:spinor-choice} and \eq{eq:anglebracket-mom}.
This is because it is then straightforward to eliminate one of the momentum
arguments by using momentum conservation and from there obtain a unique
representation of the vertex function. It is surely the case that such a
procedure would allow verification of algebraic (rather than just numerical)
equivalence for different representations of other amplitudes in the literature.

In the light of what we have learned, it is obvious that lying behind the CSW
rules is light-cone quantization. The fixed spinor $\eta$ introduced in
ref.~\cite{CSW} defines a preferred null direction which we have seen is the
same direction as the null direction $\mu$ defining light-cone time in
light-cone quantization. We note that it is the locality of the resulting
vertices in light-cone time which is actually exploited when applying the
Feynman Tree Theorem in ref.~\cite{Bill2}.

It is difficult to overstress the potential importance of the program started by
ref.~\cite{Paul}. The spinor/twistor methods have been very successful at tree
level and for cut-constructible pieces at one loop. This is the natural domain
for methods inherently tied to the structure of tree-level four-dimensional
on-shell scattering processes. Progress has been limited
beyond this domain (see however ref.~\cite{onelooprecurrence}).

However, since it transpires that these methods are a direct consequence of a
change of field variables, extending these methods simply means applying well
understood techniques from quantum field theory. For example, it is now obvious
that the CSW rules apply to fully off-shell amplitudes, the external spinors
being continued off-shell using the CSW prescription.

It also can no longer be in doubt that these methods can be appropriately
regularised (for computing general quantum corrections): we can simply apply our
favourite method to the Yang-Mills Lagrangian and trace through the consequences
for the change of variables. The real question is whether a regularisation can
be applied (hopefully dimensional regularisation appropriately adapted) in such
a way as to preserve sufficiently the simplifying power of these methods.

These regularised CSW methods will then apply to the whole amplitude to any
number of loops, excepting only that we must multiply by the correct ET matching
factors. Of course, as is well known, we cannot compute the $n$-point one-loop
all ``$+$'' amplitude or the $n$-point amplitude with just one negative helicity
leg, using only MHV rules \cite{CSW2}. These amplitudes are non-vanishing but
finite \cite{chalm} and of course contain no cut-constructible part. It is thus
a reasonable assumption that these amplitudes are missing simply because the
extra regularisation structure is missing.

Another possible extension is to develop mixed representations for computing
amplitudes, where gluonic parts can be computed in the ``MHV frame'' by
performing the relevant change of variables, while particles' interactions that
do not benefit from this (\eg involving massive particles) can be computed in
the normal ``Feynman rules frame''. It is even possible to consider mixed
representations for amplitudes containing the same field.

In \sec{ET}, we investigated the quantum contributions to the ET
matching factors, which are $Z_+=Z_-=1$ at tree level, and which
multiply S-matrix elements computed using CSW rules. These are
two-point contributions formed from the expansion coefficients
\eqs{Upsilonn}{Xijn} and the MHV-rules vertices \eq{V}. Although
Lorentz invariance has been broken by the light-cone gauge and the
change of field variables, it should be recovered when computing
S-matrix elements. Therefore the matching factors can only depend
on the external momentum squared of the scattered particle, so
providing we are dealing with only massless particles, the
one-loop and higher-loop contributions depend on no scale at all
and must be set to zero in dimensional regularisation. We verified
this argument for the divergent parts at one loop, giving also
explicit off-shell expressions for these factors. We note that in
general certain on-shell contributions to the ET matching factors
could prove to be non-zero when massive particles are included in
the theory.

\section*{Acknowledgments}
It is a pleasure to thank Andi Brandhuber, Nick Evans, Nigel
Glover, Valya Khoze, Paul Mansfield, Simon McNamara, Douglas Ross,
Olly Rosten, Bill Spence and Gabriele Travaglini for discussions,
and PPARC for financial support. Special thanks to Andi, Bill and
Gabriele for bringing to our attention ref. \cite{david},
correcting an error in an earlier version of this paper.

\appendix

\section{Off-shell matching factors}

We confirm the result \cite{david}
\be
\label{thing}
 {1\over\epsilon^2}\:g^2 C_A\,
c_\Gamma (-p^2)^{-\epsilon}
\ee
for integral \eq{kos}, where the standard one-loop factor
$c_\Gamma$ is
\[
c_\Gamma={\Gamma(1+\epsilon)\Gamma^2(1-\epsilon)
\over(4\pi)^{2-\epsilon}\Gamma(1-2\epsilon)}.
\]
Temporarily ignoring $-ig^2C_A$ and combining denominators in
\eq{kos} using Feynman parameters we have
\[
4\mu\!\cdot\!p\int\!\! {d^Dq\over(2\pi)^D}\, d\alpha\,d\beta\,
d\gamma\, \delta(1-\alpha-\beta-\gamma) {1\over\left[\alpha
q^2+\beta(p+q)^2+2\gamma\mu\!\cdot\!q\right]^3}\,.
\]
We note in passing that this has only infrared divergences.
Performing the momentum integral, the $\alpha$ integral, and
substituting $\beta=\rho(1-\omega)$ and $\gamma=\rho\omega$,
yields
\[
-2i\mu\!\cdot\!p\,
{\Gamma(1+\epsilon)\over(4\pi)^{2-\epsilon}}\int^1_0\!\!\!\!d\rho\!\int^1_0\!\!\!\!
d\omega\
\rho^{-\epsilon}(1-\omega)^{-1-\epsilon}(1-\rho\omega)^{-1+2\epsilon}
\left\{2\rho\omega\mu\!\cdot\!p-(1-\rho)p^2\right\}^{-1-\epsilon}.
\]
As noted below \eq{kos}, this integral does not in fact depend on
$\mu\!\cdot\!p$. Exploiting this \cite{david}, we set
$\mu\!\cdot\!p>0$, substitute $\omega\mapsto\omega/\mu\!\cdot\!p$
and then let $\mu\!\cdot\!p\to\infty$. The result,
\[
-2i\, {\Gamma(1+\epsilon)\over(4\pi)^{2-\epsilon}}
\int^1_0\!\!\!\!d\rho\!\int^\infty_0\!\!\!\!\!\! d\omega\
\rho^{-\epsilon} \left\{2\rho\omega
-(1-\rho)p^2\right\}^{-1-\epsilon},
\]
is readily evaluated
and gives \eq{thing} on reinstating $-ig^2C_A$.

Expanding the numerator in \eq{stuff}, the $(\mu\!\cdot\!p)^3$
term is identical to \eq{kos} and thus gives \eq{thing}. The other
terms do not have $\mu\!\cdot\!q$ on the denominator and are
therefore straightforward to evaluate. (For $p$ off-shell, they
have only ultraviolet divergences.) The result for \eq{stuff} is
therefore
\[
{1\over\epsilon^2}\:g^2 C_A\, c_\Gamma (-p^2)^{-\epsilon}+
{1\over\epsilon}\:g^2C_A\,c_\Gamma\,{11-7\epsilon\over(6-4\epsilon)(1-2\epsilon)}\,
(-p^2)^{-\epsilon}.
\]

\end{document}